\documentclass{article}
\usepackage{spconf,amsmath,graphicx}
\usepackage{graphicx}
\title{MRF Denoising with Compressed Sensing and Adaptive Filtering}
 \name{Zhe Wang$^{\star}$  \qquad Qinwei Zhang$^{\dagger}$ \qquad Jing Yuan$^{\dagger}$\qquad Xiaogang Wang$^{\star}$}
 \address{$^{\star}$ Department of Electronic Engineering, The Chinese University of Hong Kong \\
     $^{\dagger}$Department of Imaging and Interventional Radiology, The Chinese University of Hong Kong}
\begin{document}
%\ninept
%
\maketitle
\begin{abstract}
The recently proposed Magnetic Resonance Fingerprinting (MRF) technique can simultaneously estimate multiple parameters through dictionary matching. It has promising potentials in a wide range of applications. However, MRF introduces errors due to undersampling during the data acquisition process and the limit of dictionary resolution. In this paper, we investigate the error source of MRF and propose the technologies of improving the quality of MRF with compressed sensing, error prediction by decision trees, and adaptive filtering. Experimental results support our observations and show significant improvement of the proposed technologies.

\end{abstract}
\section{Introduction}
\label{sec:intro}
Magnetic Resonance Fingerprinting (MRF) recently proposed by Ma et al. [1] has drawn a lot of attentions. MRF has the potential to quantitatively examine many magnetic resonance parameters simultaneously. It has been reported that MRF outperforms the widely used DESPOT1 and DESPOT2 [2] for $T_1$ and $T_2$ estimation. It can also be used to directly estimate the combination proportions of different types of tissues at a single pixel. This may lead to new diagnostic testing methodologies.

MRF scans an object for multiple times with pseudorandomized experimental parameters, and generate unique signal evolutions (called ``fingerprints'') as a function of the multiple material properties under investigation. There is no steady state during the whole acquisition of MRF. The magnetization after the previous excitation is used as a starting point of the next excitation.
After acquisition, each pixel has a sequence of signals and the most possible material of the pixel is estimated by matching its signal evolution with a pre-calculated dictionary, which includes the ``fingerprints'' of possible materials.  The dictionary entry that has  the maximum dot-product with the signal evolution is considered to be most likely to represent the true signal evolution. Each dictionary entry is also associated with a set of magnetic parameters. Quantitative maps of the magnetic parameters  are then translated from the results of dictionary matching in a pixel-wise fashion. Since MRF is very new, little further work has been done to improve the quality of its estmation on parameter maps.

The contribution of this paper lies in two aspects. Firstly, we investigate the error sources of MRF on estimating parameter maps, and have the following observations.
\begin{itemize}
	\item In order to achieve the tradeoff between accuracy and scanning time, MRF heavily undersamples data in k-space and leads to artifacts in the reconstructed image at each sampling time. Such errors propagate to parameter maps through dictionary matching.
	\item Because of the limit of memory and computational resource, the resolution of the dictionary is limited. The dictionary size grows exponentially with the number of parameters to be estimated. In our experiments, although only three parameters are estimated, the dictionary size has reached $100$ million. We observe that most of the errors happen when a true parameter lies between dictionary entries, because the similarities between the measured signal evolution and all the dictionary entries are relatively low in that case. Some statistic results are shown in Figure 1.
	\item Since MRF estimates parameters by dictionary matching, if a dictionary entry is wrongly matched, its parameter could be far away from the ground truth, which leads to large errors. An example is shown in Figure 2. The estimation errors are not Gaussian distributed.
\end{itemize}

% Below is an example of how to insert images. Delete the ``\vspace'' line,
% uncomment the preceding line ``\centerline...'' and replace ``imageX.ps''
% with a suitable PostScript file name.
% -------------------------------------------------------------------------

\begin{figure}[t]
\centering
 \begin{minipage}{0.4\linewidth}
   \centering
   \includegraphics[width=1\linewidth]{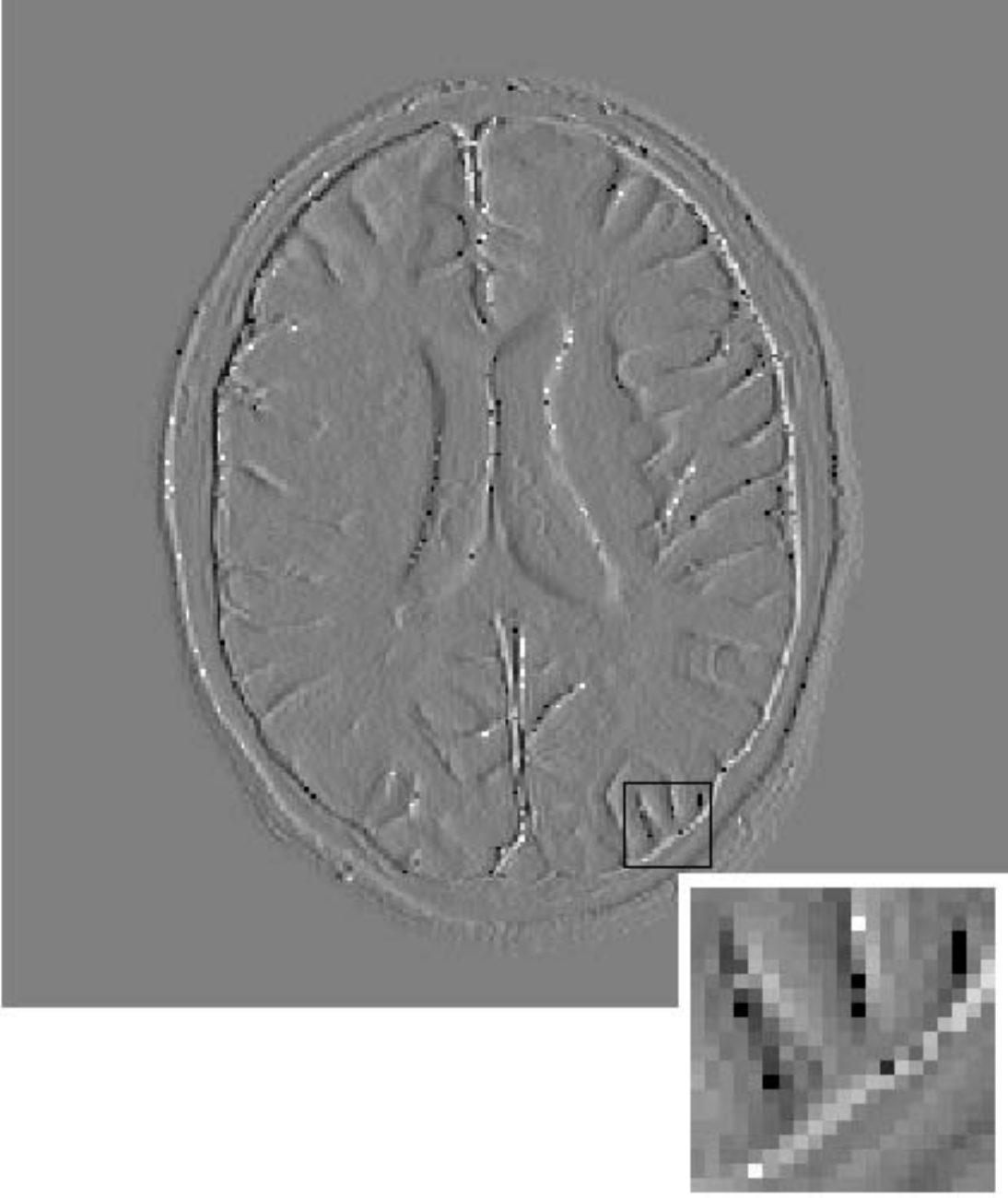} \\ (a)
 \end{minipage}
 \begin{minipage}{0.58\linewidth}
   \centering
   \includegraphics[width=1\linewidth]{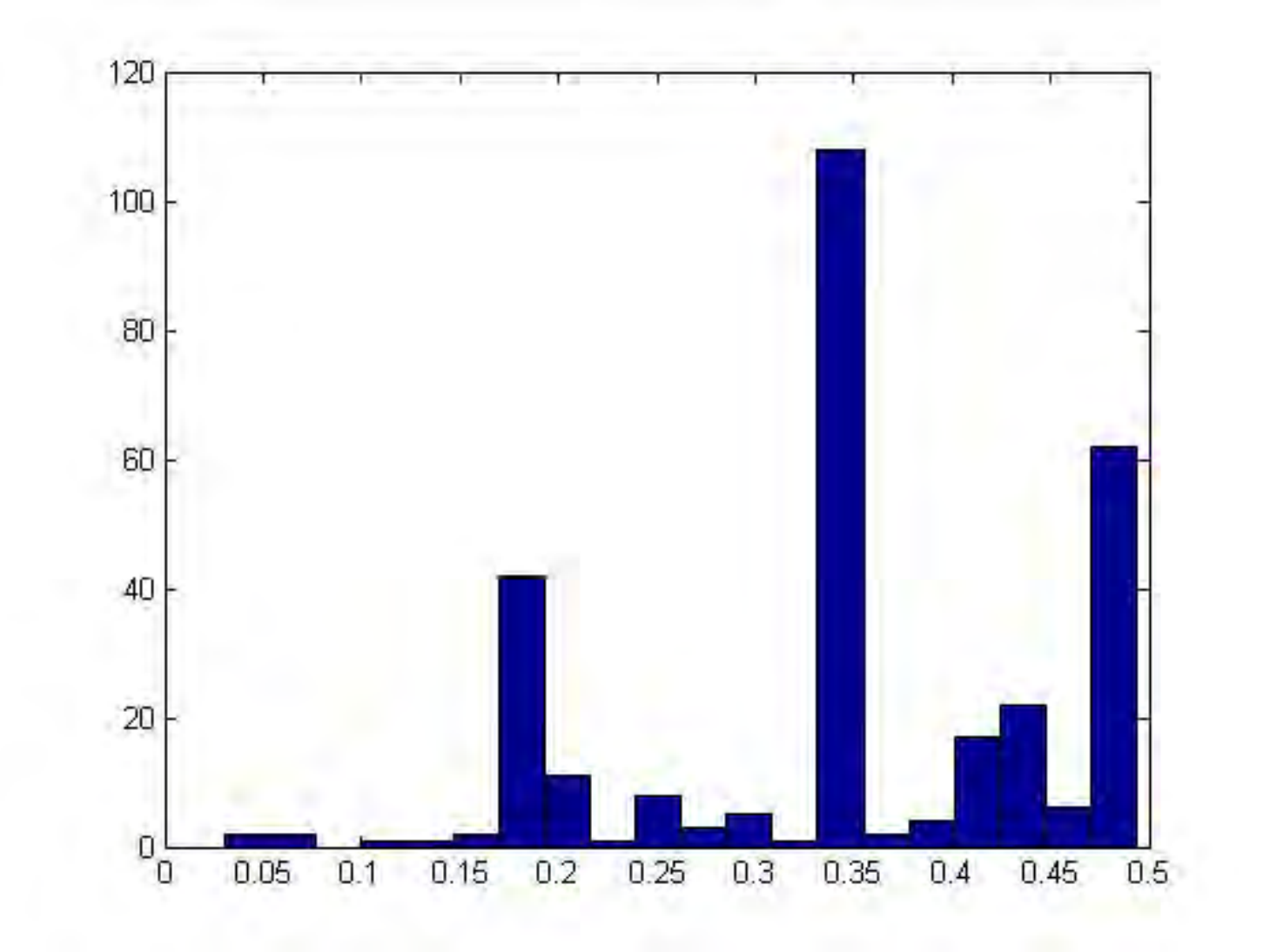} \\ (b)
 \end{minipage}
  \begin{minipage}{1\linewidth}
   \centering
   \includegraphics[width=1\linewidth]{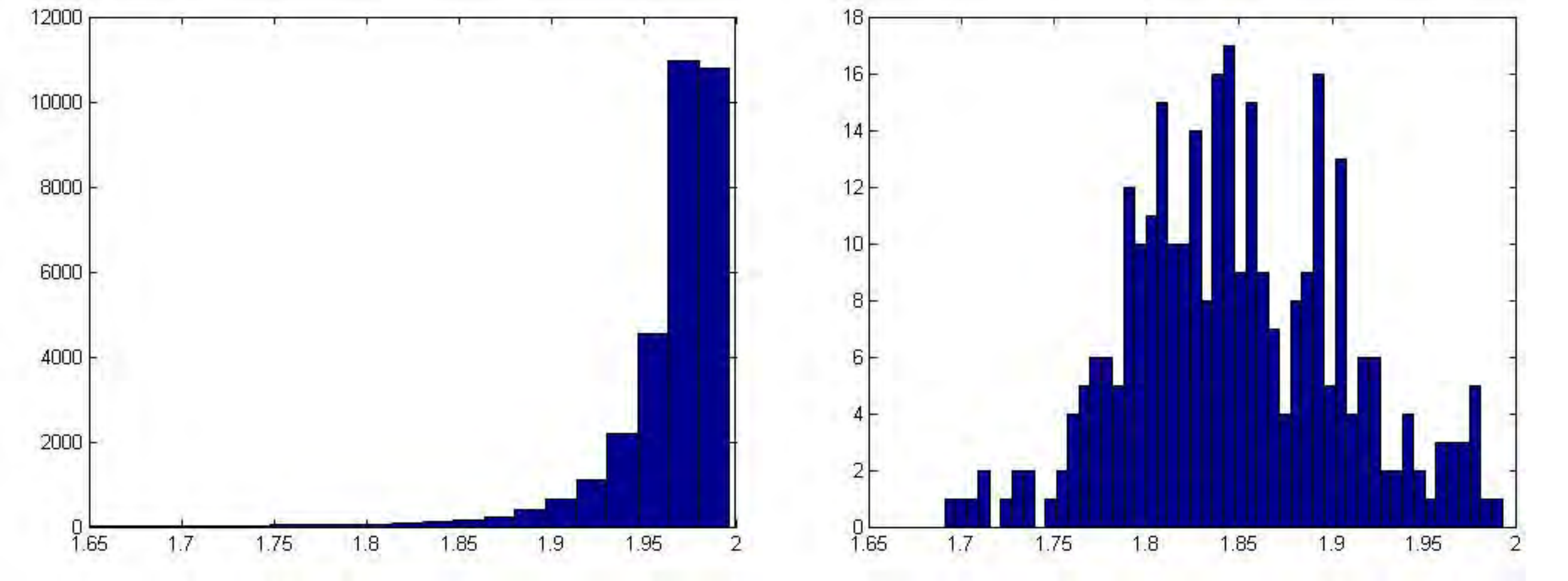}
   \centerline{(c) correct pixels~~~~~~~~~~~~~~~~~~~~(d) incorrect pixels}\medskip
 \end{minipage}
\caption{(a) Estimated off-resonance frequency map by MRF. Some pixels have large errors because of mismatch with dictionary entries. (b) Histogram of the distance between the ground truth parameter and its nearest dictionary entry for all the mismatched pixels. The distance between two neighbour entries is taken as 1. Most mismatched pixels lies in the middle of dictionary entries. (c)-(d): Histograms of the similarities between the signal evolutions and the best matched dictionary entries. For correctly matched pixels, the similarities are mostly above 1.9, while those of mismatched pixels span between 1.8 and 1.9. The statistics are obtained from the train set in Figure \ref{fig:TheBrain_result}.}
\label{fig:fingerprint}
\end{figure}

Secondly, based on the observations above, we propose three technologies to correct estimation errors. (1) We apply Compressed Sensing (CS) [3] at each sampling time point, and reconstruct all the k-space data before the matching process. (2) A method is proposed to predict the correctness of every pixel on the parameter map with decision trees [4]. The prediction is based on the top matching similarities and corresponding indices among all the dictionary entries. (3) If a pixel is predicted as error, it is replaced with the result of convolving its neighbor pixels with an adaptive filter. It does not include the value of the error pixel in the linear combination, since its error could be very large. The weights on the neighbor pixels are adaptively decided. %It is effective on removing isolated error pixels and preserve important image structures in the meanwhile.

\section{Methods}
\label{sec:methods}
The key assumption underlying MRF is that the signal evolutions or fingerprints for different materials or tissues can be generated  with an appropriate acquisition scheme: $x^i = f(para_1,para_2,para_3,exp_1^i,exp_2^i)$, $i = 1,2,...N$, where $N$ is the number of sampling times. $x^i$ is the signal evolution at the $i$th sampling time. $para_1$, $para_2$, and $para_3$ are the parameters of the material to be estimated. $exp_1^i$ and $exp_2^i$ are the chosen experimental parameters at the $i$th sampling time. $f$ is the physical model of generating the signal evolution, derived from the well-known Bloch equation formalism of magnetic resonance. In this study, we choose $f$ as an inversion-recovery balanced steady state free-precession(IR-bSSFP) sequence suggested in [1]. $para_1$, $para_2$ and $para_3$ are the longitude relaxation time $T_1$, the transverse relaxation time $T_2$, and the off-resonance frequency $df$. $exp_1$ and $exp_2$ are the flip angle FA and the repetition time TR.

Dictionary $D \in \mathcal{C}^{N\times K}$ is a collection of signal evolutions for possible combinations of materials given the same set of experimental parameters. $K$ is the size of the dictionary, \textit{i.e.} the total number of possible combinations. The goal of MRF is to find the right combination from the dictionary, which is most likely to be  the observed signal evolution $\hat{x}\in \mathcal{C}^{N\times 1}$,
\begin{equation}
[T_1,T_2,df] = g(D,\hat{x}).
\end{equation}
$g$ is the function to select an entry from the dictionary best matching $\hat{x}$. In [1], the dot-product is used to calculate the similarity between the measured signal evolution and all dictionary entries $\{D_k\}$,
\begin{equation}
g(D,\hat{x}) = \Gamma(\max_{k}\{D_k^{H} \hat{x} + \hat{x}^{H} D_k \}). \label{eq:sim}
\end{equation}
Both $D_k$ and $\hat{x}$ are complex signals with unit $L_2$ norm ($||D_{k}||^2 = ||\hat{x}||^2$=1). $D_k^{H}$ and $\hat{x}^{H}$ denotes the conjugate transpose of $D_k$ and $\hat{x}$. $\Gamma$ is the mapping from a dictionary index to the corresponding parameters $T_1$, $T_2$ and $df$.

\begin{figure}[t]
\begin{minipage}{1.0\linewidth}
%  \centering%
  \centerline{\includegraphics[width=8.5cm]{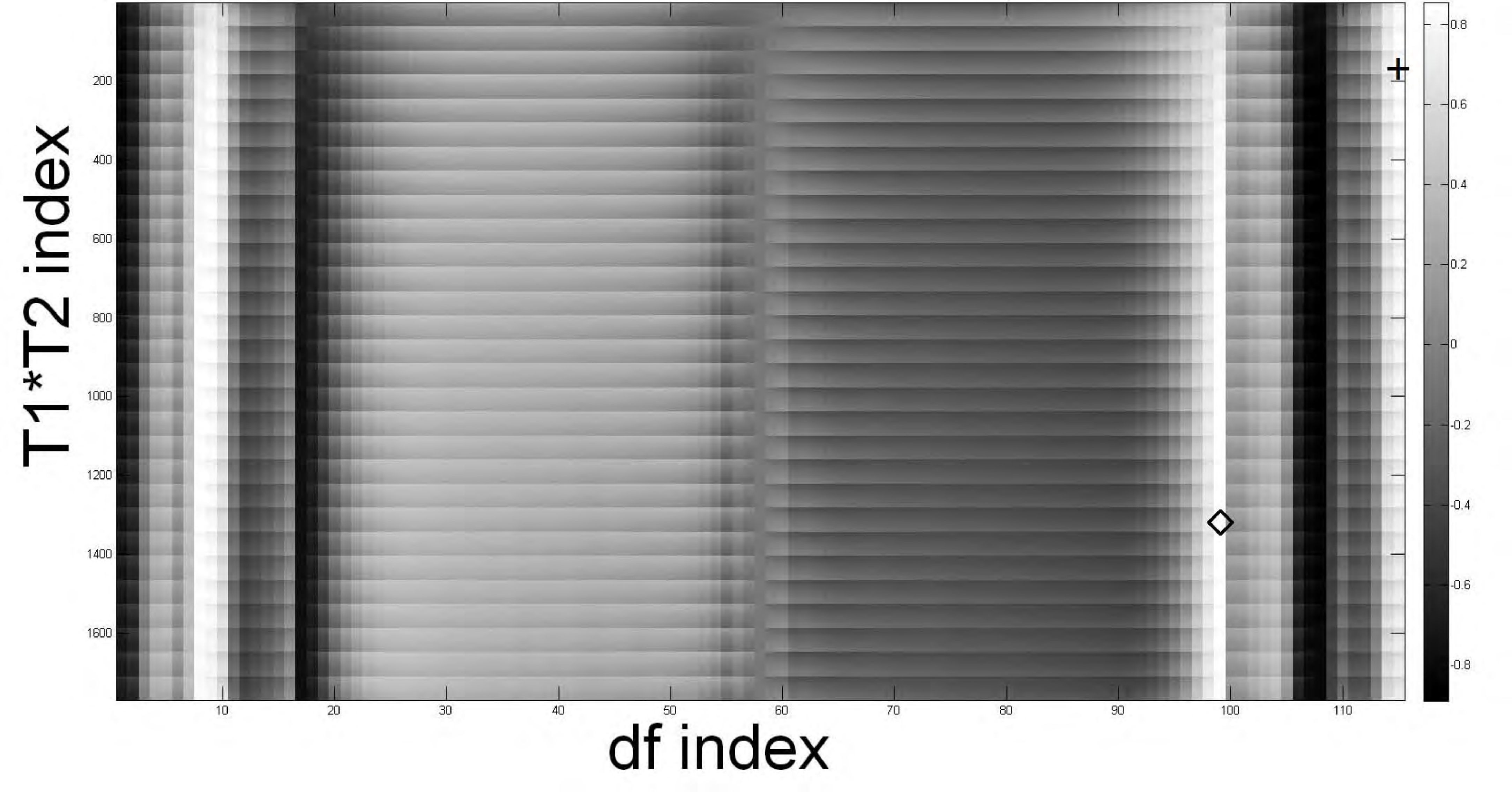}}
%  \vspace{2.0cm}
%  \centerline{}\medskip
\end{minipage}
\caption{The similarity vector for a pixel lying between two entries. Multiple local maximum which are far apart are observed. They lead to large estimation error. The entry with the highest similarity is marked with ``+'', and the entry nearest to the ground truth is marked with ``$\diamond$''.}
\label{fig:similarity_vector}
\end{figure}

Let $s_k = D_k^{H} \hat{x} + D_k \hat{x}^{H}$ be the similarity between the measured signal evolution and a dictionary entry. $\hat{x}$ is contaminated with various noise. According to our experiments, if the ground truth parameter lies in the middle of two entries, it is likely that there are multiple local maximums with similar similarities. These local maximums may be far apart in the dictionary, which leads to mismatches and large estimation errors. An example is shown in Fig. \ref{fig:similarity_vector}.

The observations motivate us to improve the quality of MRF in two stages. In the first stage, we apply CS [3] to the undersampled k-space data at each sampling time point for image reconstruction. A sequence of $N$ images is reconstructed. The artifacts caused by undersampling is largely removed by incorporating the sparsity prior in the wavelet domain and finite difference domain. %It is shown by our experiments that applying CS alone before the matching process is able to significantly improve the quality of parameter maps.

In the dictionary matching stage, to cope with the errors caused by dictionary resolution, we propose to use pre-trained decision trees to predict the locations of wrongly matched pixels, and then replace them with the results of convolving their neighbor pixels with a adaptive filter. If a pixel contains material whose parameters, especially the parameter of off-resonance frequency, are not close to any dictionary entries, not only the similarity between the measured signal evolution and the best matched dictionary entry is relatively low, but also there are multiple local maximums far apart in the dictionary. On the contrary, if a pixel is correctly matched, its largest similarity in the dictionary is high and also much higher than other peaks. Some statistics are shown in Figure \ref{fig:fingerprint}. Therefore, the values of the local maximums and their locations are all useful cues of predicting mismatched pixels.

Three decision trees will be trained for $T_1$, $T_2$ and off-resonance respectively. A decision tree can determine the class label (correctly or wrongly matched pixel in our case) of a pixel from the values of its input attributes. 
Each of its internal nodes represents a test on an attribute. Each branch represents the outcome of the test. Each leaf node is associated with a class label and represents the decision taken after computing all attributes. A path from root to leaf represents classification rules. 
The input attributes of these decision trees form an $8$ dimensional feature vector. It includes the largest similarity values of the four highest peaks and their indices. To train the decision trees, we simulated $T_1$, $T_2$ and off-resonance frequency maps for a MR image. The ground truth of the parameters on the simulated data is known, and therefore we know whether pixel is correctly matched or not.

\begin{figure}[t]
\begin{minipage}{1.0\linewidth}
%  \centering%
  \centerline{\includegraphics[width=\linewidth]{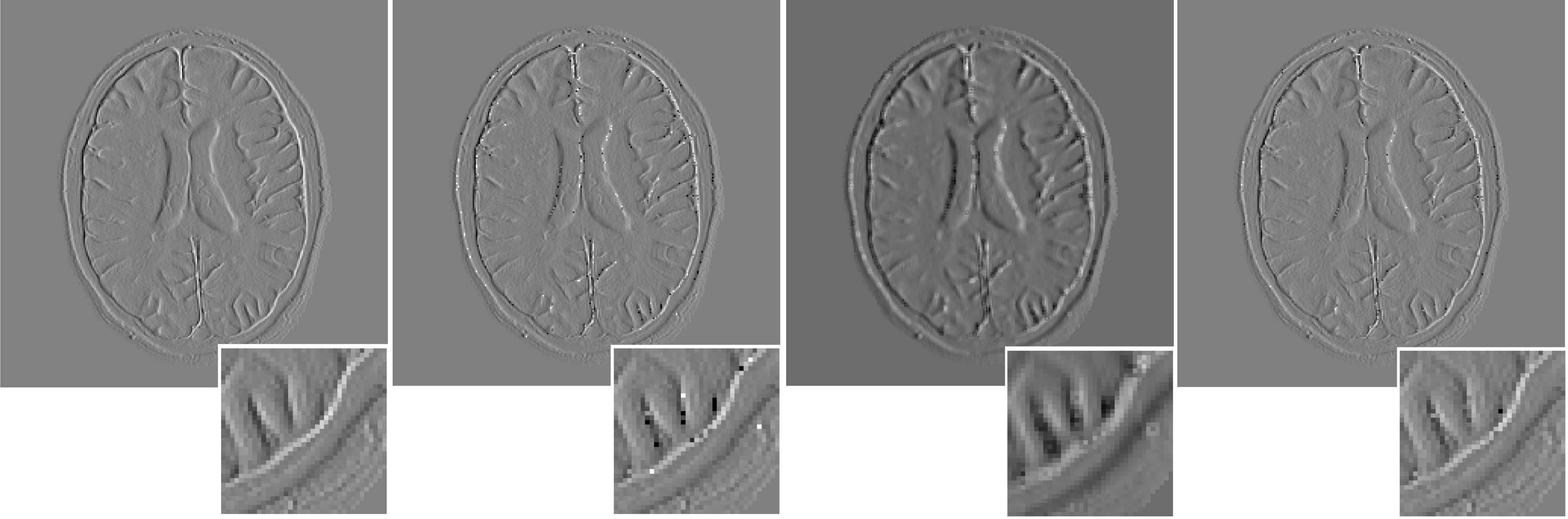}}
%  \vspace{2.0cm}
%  \centerline{}\medskip
\end{minipage}
\caption{Comparison of filtering results. From left to right: ground truth, CS+MRF, Gaussian filtering on the result of CS+MRF, and our adaptive filtering.}
\label{fig:filter}
\end{figure}

By predicting the correctness of pixels with the decision trees, we divide all the pixels into two classes. A pixel predicted as being correctly matched follows the original MRF matching process. Otherwise,  it is replaced by convolving its neighbor pixels with an adaptive filter, defined as:
\begin{eqnarray}
I_{filtered}(p) = \frac{1}{W}\sum_{j\in \Omega/p}w_{p_j} I_{p_j}. \label{eq:filter}
\end{eqnarray}
Since the centered pixel $p$ is mismatched, it may have large error. We exclude it from the combination in Eq (\ref{eq:filter}). $\Omega$ is the neighborhood of $p$. $I_{p_j}$ is the estimated parameter at pixel $p_j$. $w_{p_j}$ is adaptively computed with Eq (\ref{eq:weight}). $W$ is a normalization factor to ensure the weights to sum to 1.
\begin{eqnarray}
w_{p_j} = exp\left(-\frac{\Vert p- p_j\Vert^2_2}{\sigma_d^2}\right)exp\left(-\frac{\Vert 2- s_{p_j}\Vert^2_2}{\sigma_s^2}\right). \label{eq:weight}
\end{eqnarray}
The first term in Eq (\ref{eq:weight}) is a spatial Gaussian weighting that decreases the influence of distant pixels. The second term is a similarity Gaussian weighting that decreases the influence of the pixels whose dictionary matching result is unreliable. $s_{p_j}$ is the largest similarity of dictionary matching at pixel $p_j$. According to Eq (\ref{eq:sim}), its maximum value is 2. Figure \ref{fig:filter} compares the result of our adaptive filtering and commonly used Gaussian filtering. Our approach can well preserve local structures with less blurring effect.

\section{result}
\label{sec:result}
\subsection{Experimental Settings}
\label{ssec:sqdict}
All the experiments are implemented with Matlab 2012b(The MathWorks). We simulate the MRF signal evolution with the IR-bSSFP sequence using a pseudorandomized series of flip angles and repetition time uniformly sampled between 10.5ms and 14ms. The flip angles are calculated as a series of repeating sinusoidal curves added by Gaussian random noise. The total number of sampling points is $500$. When the sampling time $0\leq t\leq 250$, $\text{FA}(t) = 10+\sin(\frac{2\pi}{500}t)\times 50+\eta$, where $\eta$ is a noise term sampled from a Gaussian distribution with a standard deviation of $5$. When $251\leq t \leq 300$, $\text{FA}(t) = 0$. When $301 \leq t \leq 500$, $\text{FA}(t) = 5+\sin(\frac{2\pi}{200}t)\times 25+\eta$. This simulation is the same as described in [1].

\begin{figure}[t]
\begin{minipage}{1.0\linewidth}
%  \centering
  \centerline{\includegraphics[width=8.5cm]{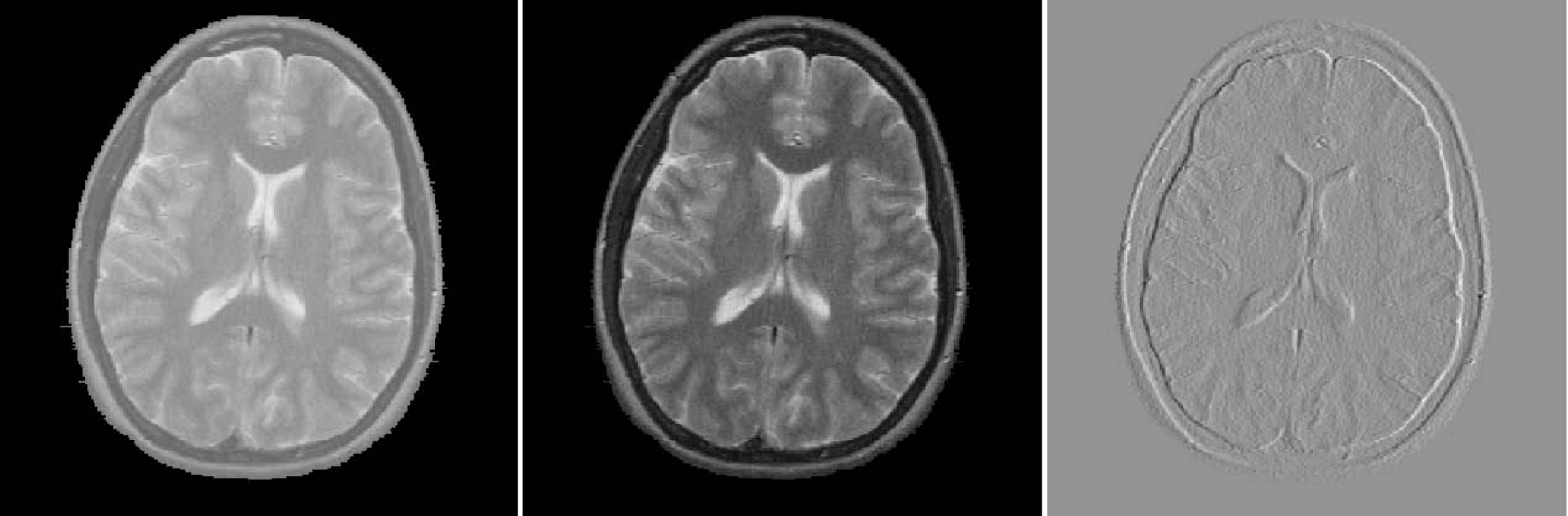}}
%  \vspace{2.0cm}
\end{minipage}%
\caption{The set of parameter maps used for training the decision trees. From left to right: $T_1$, $T_2$ and off-resonance frequency maps}
\label{fig:simu_train}
\end{figure}

All the $T_1$, $T_2$ and off-resonance frequency maps are simulated with ground truth. A set of $T_1$, $T_2$ and off-resonance frequency maps are used for training decision trees as shown in Figure \ref{fig:simu_train}. Another two sets of maps are simulated for test. Because of space limit, we only show one test set in Figure \ref{fig:TheBrain_result}. The size of the designed dictionary  $61 \times 29 \times 115\times 500$. The entries for $T_1$ are uniformly sampled from 800ms to 2000ms every 20ms. The entries of $T_2$ are uniformly sampled from 20ms to 300ms every 5ms. The sampling rates of the entries of off-resonance frequency are 2Hz in [-80Hz 80Hz], 10Hz in [80Hz 250Hz] and [-250Hz -80Hz] in order to incorporate the effect of signal evolutions in different $B_0$ fields. This design was also used in [1].
The k-space data at each sampling time point was undersampled by a variable density random mask. The undersampling ratio was chosen to be 70\%. The same mask was applied to both the training images and the test images.

\subsection{Estimation of parameter maps}
\label{ssec:compare}
\begin{figure}[t]%
\centering
\begin{minipage}[b]{0.8\linewidth}
  %\centering
  \centerline{\includegraphics[width=8.1cm]{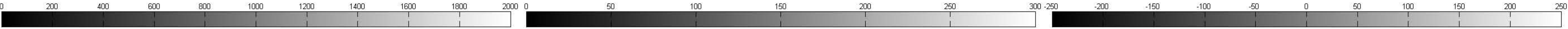}}
%  \vspace{2.0cm}
\end{minipage}

\begin{minipage}[b]{0.8\linewidth}
%  \centering
  \centerline{\includegraphics[width=8.1cm]{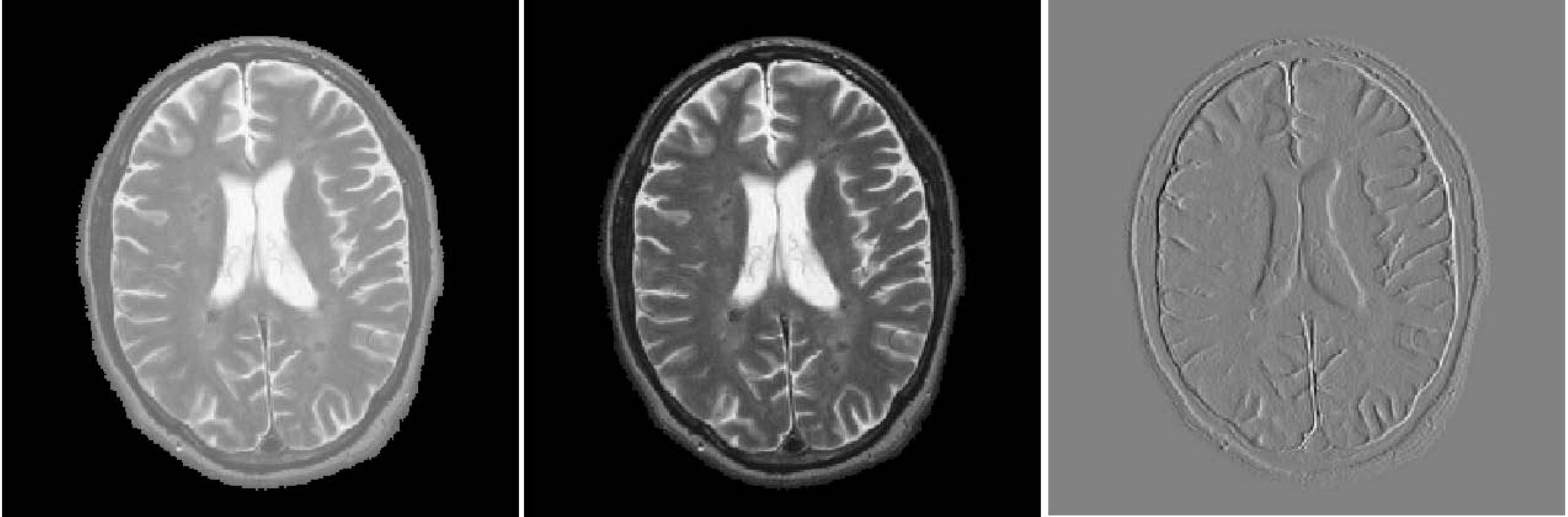}}
%  \vspace{2.0cm}
\end{minipage}

\begin{minipage}[b]{0.85\linewidth}
%  \centering
  \centerline{\includegraphics[width=8.1cm]{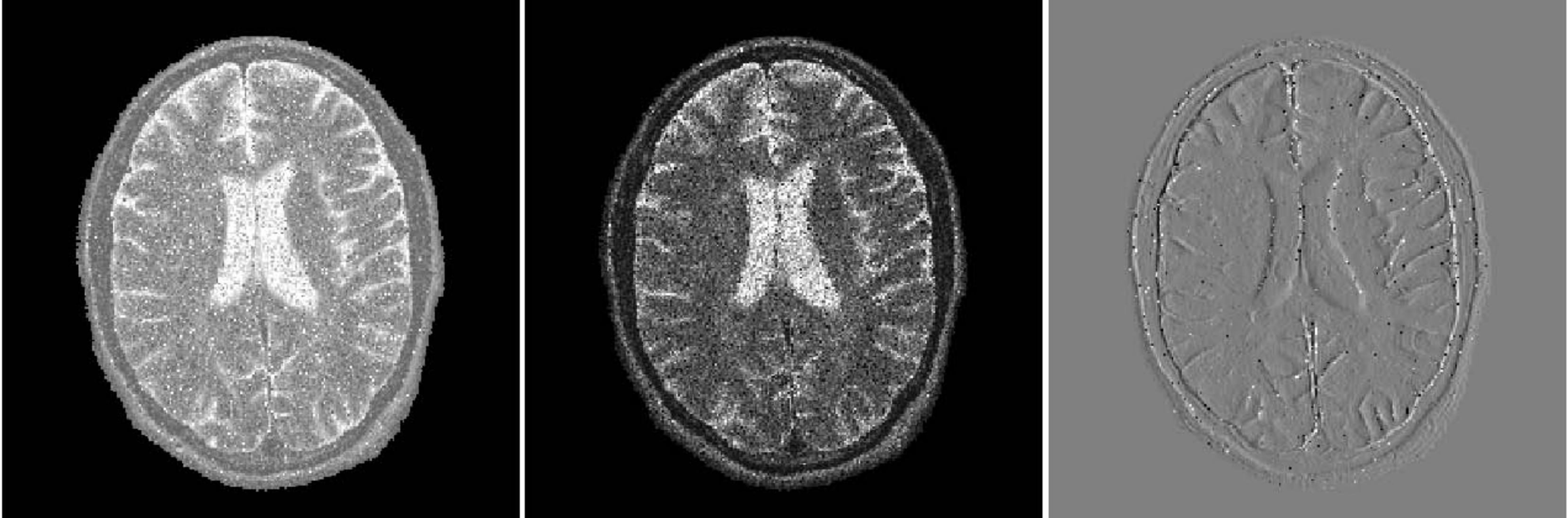}}
%  \vspace{2.0cm}
\end{minipage}

\begin{minipage}[b]{0.8\linewidth}
%  \centering
  \centerline{\includegraphics[width=8.1cm]{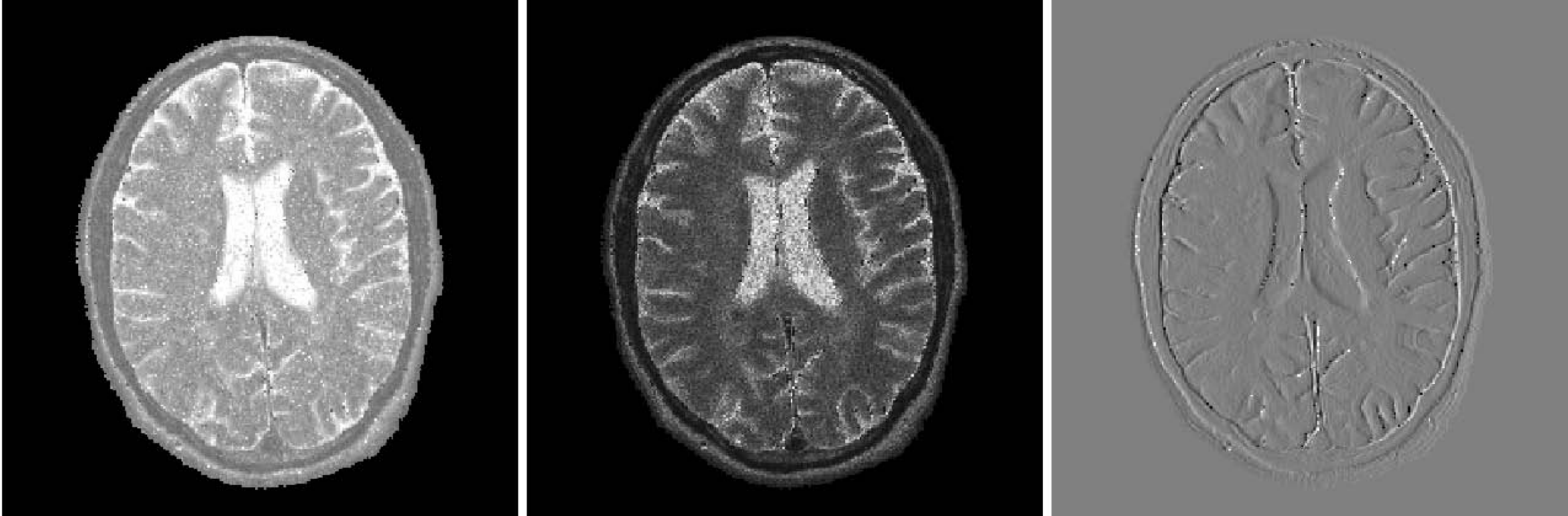}}
%  \vspace{2.0cm}
\end{minipage}

\begin{minipage}[b]{0.8\linewidth}
%  \centering
  \centerline{\includegraphics[width=8.1cm]{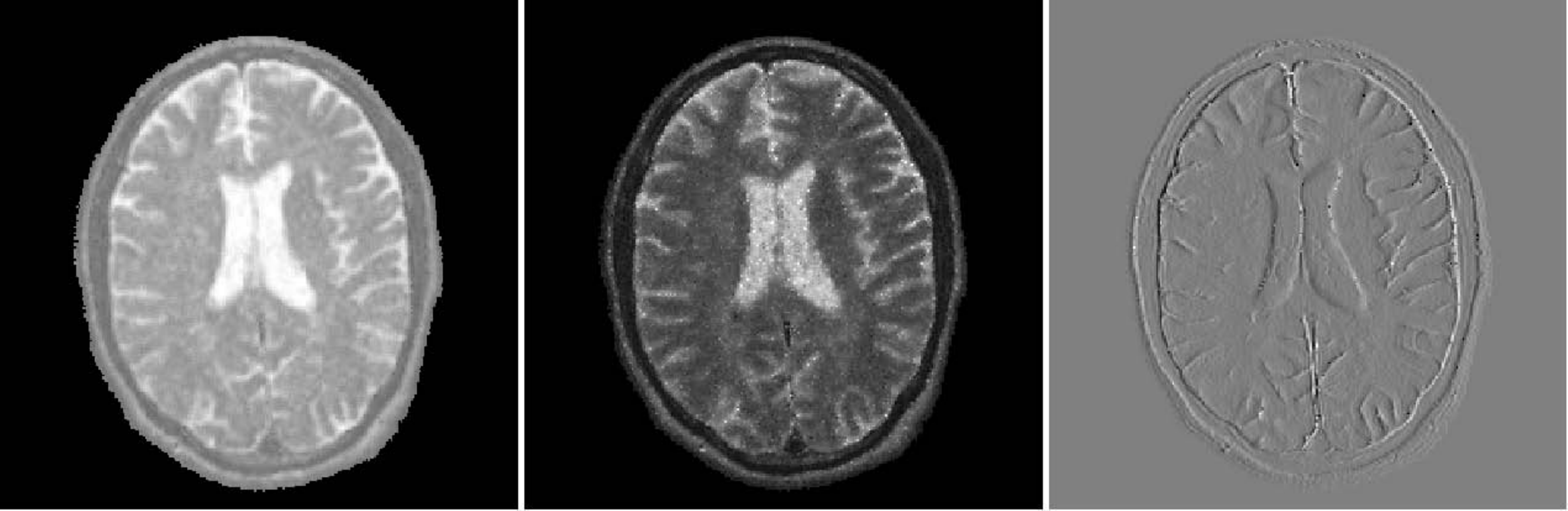}}
%  \vspace{2.0cm}
\end{minipage}
\small\caption{Results of MRF, CS-MRF and CS-Tree8AF. From left to right: $T_1$, $T_2$ and off-resonance frequency. From top to bottom: ground truth, MRF, CS, and CS-Tree8AF.}
\label{fig:TheBrain_result}
\vspace{-0.2cm}
\end{figure}

Fig.\ref{fig:TheBrain_result} shows the estimated parameter maps for a test set. It clearly shows the effectiveness of compressed sensing and adaptive filtering on removing noise generated by MRF. Table \ref{table_psnr} reports the quantitative comparison with the measurements of Peak Signal to Noise Ratio(PSNR) and Structural SIMilarity(SSIM) index [5]. The evaluation is done on two test sets of parameter maps. Our proposed approach (CS-Tree8AF) has the best performance under both measurements and significantly improve the quality of parameter maps estimated with MRF. In order to evaluate the effectiveness of each component in our approach, we compare with a few alternatives. (1) Only apply compressed sensing as preprocessing of MRF however without adaptive filtering (CS). It clearly improves MRF, but not as good as CS-Tree8AF. (2) Different than our decision trees with $8$ attributes as input, it only has the largest similarity after dictionary matching as the input of decision trees (CS-Tree1AF). It essentially predicts the mismatched pixels by simple thresholding. (3) The decision trees have four input attributes, i.e. the largest similarity values of the four highest peaks after dictionary matching (CS-Tree4AF). However, it ignores the indices of the peaks. CS-Tree1AF and CS-Tree4AF are not as good as CS-Tree8AF, since their decision trees cannot accurately predict mismatched pixels.

\begin{table}
\begin{tabular}{|c|c|c|c|}
\hline
  & $T_1$ & $T_2$ & df  \\
\hline
MRF & 24.3$\setminus$ 0.731 & 20.4$\setminus$  0.706& 22.9$\setminus$ 0.867  \\
\hline
CS & 26.2 $\setminus$0.801 & 21.5$\setminus$ 0.802& 26.4$\setminus$  0.961  \\
\hline
%CS-AF & 28.1 $\setminus$ 0.875 & 25.3 $\setminus$ 0.863  & 21.6 $\setminus$ 0.700 \\
%\hline
CS-Tree1AF & 27.1 $\setminus$ 0.822 & 22.4 $\setminus$ 0.818 & 27.2 $\setminus$ 0.955 \\
\hline
CS-Tree4AF & 28.0 $\setminus$ 0.873 & 22.3 $\setminus$ 0.842 & 26.5 $\setminus$ 0.955 \\
\hline
CS-Tree8AF & 28.4 $\setminus$ 0.881& 22.9$\setminus$ 0.863 & 28.9 $\setminus$ 0.974 \\
\hline
\end{tabular}
\caption{PSNR and SSIM comparison of MRF, CS, CS-Tree1AF, CS-Tree4AF and CS-Tree8AF. CS-Tree8AF is the final approach we proposed. Others are used to evaluate the effectiveness of each component in our approach. In each cell, the left number denotes PSNR in dB and the right number denotes SSIM. See details in the text of Section \ref{ssec:compare}.}
\vspace{-0.5cm}
\label{table_psnr}
\end{table}
\section{Discussion and Conclusion}
MRF has great potentials of developing new diagnostic testing methodologies. It is important to understand its error sources and improve its quality on parameter estimation. In this paper, we investigate two types of errors through empirical study. Motivated by our empirical observations, the technologies of compressed sensing, error prediction, and adaptive filtering are proposed to improve the MRF quality. Their effectiveness is shown through experiments. Dictionary learning based CS, more attributes for decision trees and patch-based filters can be further explored in the future.

\label{sec:conclusion}

\small\section{Acknowledgment}
This work was supported in part by Hong Kong RGC grant SEG $\_$CUHK02, CUHK418811, China NSFC grant 81201076.
%\vfill
%\pagebreak

\bibliographystyle{IEEEbib}
\vspace{-0.2cm}
%\small\bibliography{strings,refs}

\section{REFERENCE}
\begin{minipage}{0.02\textwidth}
[1] 
\end{minipage}
\begin{minipage}{0.001\textwidth}
 
\end{minipage}
\begin{minipage}[t]{0.46\textwidth}
Dan Ma, Vikas Gulani, Nicole Seiberlich, Kecheng Liu, Jeffrey
L Sunshine, Jeffrey L Duerk, and Mark A Griswold, “Magnetic
resonance fingerprinting,” Nature, vol. 495, no. 7440, pp.
187–192, 2013.
\end{minipage}

\vspace{0.01\textwidth}
\noindent \begin{minipage}{0.02\textwidth}
[2]
\end{minipage}
\begin{minipage}{0.001\textwidth}
 
\end{minipage}
\begin{minipage}[t]{0.46\textwidth}
Sean CL Deoni, Terry M Peters, and Brian K Rutt, “High resolution
t1 and t2 mapping of the brain in a clinically acceptable
time with despot1 and despot2,” Magnetic resonance in
medicine, vol. 53, no. 1, pp. 237–241, 2005.
\end{minipage}

\vspace{0.01\textwidth}
\noindent \begin{minipage}{0.02\textwidth}
[3]
\end{minipage}
\begin{minipage}{0.001\textwidth}
 
\end{minipage}
\begin{minipage}[t]{0.46\textwidth}
Michael Lustig, David Donoho, and JohnMPauly, “Sparse mri:
The application of compressed sensing for rapid mr imaging,”
Magnetic resonance in medicine, vol. 58, no. 6, pp. 1182–1195,
2007.
\end{minipage}

\vspace{0.01\textwidth}
\noindent \begin{minipage}{0.02\textwidth}
[4]
\end{minipage}
\begin{minipage}{0.001\textwidth}
 
\end{minipage}
\begin{minipage}[t]{0.46\textwidth}
J. Ross Quinlan, “Induction of decision trees,” Machine learning,
vol. 1, no. 1, pp. 81–106, 1986.
\end{minipage}

\vspace{0.01\textwidth}
\noindent \begin{minipage}{0.02\textwidth}
[5]
\end{minipage}
\begin{minipage}{0.001\textwidth}
 
\end{minipage}
\begin{minipage}[t]{0.46\textwidth}
Zhou Wang, Alan C Bovik, Hamid R Sheikh, and Eero P Simoncelli,
“Image quality assessment: From error visibility to
structural similarity,” Image Processing, IEEE Transactions on,
vol. 13, no. 4, pp. 600–612, 2004.
\end{minipage}

\end{document}